# Optically Actuated Transitions in Multimodal, Bistable Micromechanical Oscillators

Lior Michaeli[†], Ramon Gao[†], Michael D. Kelzenberg, Claudio U. Hail, John E. Sader, and Harry A. Atwater[*]

Department of Applied Physics and Materials Science, California Institute of Technology, Pasadena, CA 91125, USA

[†]These authors contributed equally: Lior Michaeli, Ramon Gao

[*] *haa@caltech.edu*

## Abstract

We experimentally demonstrate a new class of optomechanical nonlinearities in weakly damped micromechanical resonators, arising from the interplay between the Duffing nonlinearity, intermodal coupling, and thermal fluctuations. Within the bistable regime of a single Duffing mode driven by radiation pressure forces, we observe stochastically generated sidebands, originating from thermal fluctuations around equilibrium trajectories in phase space, and exploit these sidebands to induce probabilistic transitions between bistable states using weak secondary acoustic excitation. Extending this framework to multimodal interactions, we show that nonlinear modes coupling within the same resonator leads to similar transitions due to parametric modulation around the noise-excited sidebands as a result of frequency mixing. Simultaneously, abrupt changes in displacements of modes cause their instantaneous energy exchange rates to span five orders of magnitude. These findings open new avenues for reconfigurable optomechanical networks, nonreciprocal energy transport, and precision sensing based on dynamically tunable mechanical nonlinearities.

## Introduction

Micromechanical resonators provide a powerful platform for exploring nonlinear and stochastic dynamics at the mesoscopic scale. Their nonlinearities and low dissipation have enabled observation of classical and quantum noise squeezing[1–5], instability and multistability of vibrational states[6–9], synchronization[10,11], chaotic dynamics[12–14], and intermodal energy exchange[15–17]. These unique properties have also been harnessed to enhance the sensitivity of micro-mechanical resonators in force and mass detection[8,18–20], and in precision metrology[21,22], highlighting their significance for both fundamental research and practical applications.

Weakly damped resonators with Duffing-type nonlinearities exhibit noise-induced sidebands in their spectral response[23]. These sidebands have recently attracted significant experimental interest by offering a means to read out or control the resonator state[2,3,14,24,25], providing spectral signatures of noise squeezing[2,3,24], enabling generation of frequency combs from a single mechanical mode[25], and offering a systematic approach to chaos generation[14]. Additionally, the intrinsic bistability of Duffing-type modes motivated extensive theoretical studies of noise-induced escape from effective potential wells under varying drive and noise conditions[26–32].

Beyond single-mode dynamics, there is growing interest in leveraging multiple mechanical modes[16,33–42], where their tailorable interactions unlock new possibilities for coherent phonon manipulation[33,34], sensing[35], frequency stabilization[16,36], exotic nonreciprocal states[37–40] and exploration of new dynamical phases of matter[41,42]. These interactions can be mediated by external stimuli, e.g., optical or electrical fields[37–42], or emerge intrinsically through nonlinear intermodal coupling, recently shown to host a rich variety of physical phenomena[9,15,17,19,43–45]. In this regime, the orthogonality of a resonator's vibrational modes is broken, and one mode's resonance frequency depends on another mode's amplitude. Despite significant progress, the interplay between nonlinear modal interactions and noise-driven processes remains largely unexplored. Unraveling this connection could unlock new strategies for controlling vibrational states, tuning energy transfer, and engineering novel nonlinear phenomena in micromechanical resonators.

Here, we demonstrate optomechanical nonlinear phenomena in weakly damped mechanical resonators arising from the interplay between the Duffing nonlinearity, intermodal coupling, and stochastic fluctuations. We use radiation pressure forces from an intensity-modulated laser to drive the resonator's fundamental mode into bistability characterized by Duffing nonlinearity. We then show that introducing a weak secondary excitation detuned from the main excitation by the low-frequency stochastic sideband injects energy into the slow vibrational dynamics, triggering transitions from high- to low-amplitude states. To quantify the role of thermomechanical stochastic noise in these transitions, we measure the transition probability versus secondary excitation frequency and amplitude. Transition efficiency is maximized either by increasing the perturbation amplitude or around the sideband frequencies like the parametric instabilities associated with Arnold tongues[46,10].

As micromechanical resonators naturally support multiple vibrational modes, we extend our framework to a multimodal system, with intrinsic nonlinear coupling arising from displacement-induced tension. By applying an acoustic secondary excitation near the higher-order mode, we induce amplitude and eigenfrequency transitions across the two considered modes. Beating between the probe and the noise-driven higher-order mode induces parametric excitation of the fundamental mode's sideband, enabling the system to transition to the lower-amplitude state. Owing to the fundamental mode's large dynamic range, these amplitude transitions span from microns to nanometers. Accordingly, the intermodal coupling strength shifts across five orders of magnitude.

## Results and Discussion

Micromechanical resonators driven at low amplitudes are well described by a driven damped harmonic oscillator. However, nonlinear effects become significant with increasing amplitude, giving rise to more complex dynamical behavior. These nonlinearities may be dispersive, modifying the resonance frequency, or dissipative, affecting the energy dissipation rate. In particular, a cubic term in the restoring force, typically arising from geometric nonlinearities for flexural structures, leads to amplitude-dependent resonance frequency and bistability[47].

Our device is a 50 nm-thick silicon nitride trampoline resonator, comprising a central membrane suspended by compliant serpentine springs (Fig. 1, inset). This geometry offers both high susceptibility and a wide dynamic range, with nonlinearities emerging only at large vibrational amplitudes. The dynamics of the fundamental flexural mode in response to an external driving force $F$ can be described by the following Duffing equation of motion (EOM) in terms of the displacement amplitude $q$:

$$\ddot{q} + 2\Gamma\dot{q} + \omega_0^2 q + \alpha q^3 + \eta \dot{q} q^2 = F/m \quad (1)$$

Where $\Gamma = 0.37$ Hz is the linear damping rate, $f_0 = \omega_0/2\pi = 5{,}862$ Hz the resonance frequency, $m = 1.83 \times 10^{-13}$ kg the effective mass, $m \cdot \alpha = 1.83 \times 10^6$ N/m$^3$ the Duffing nonlinearity and $m \cdot \eta = 2.13 \times 10^{-1}$ N·s/m$^3$ the nonlinear damping rate. Here, the resonator is actuated by radiation pressure force $F$ proportional to the laser beam power $P$. Laser intensity modulation at frequency $\omega_d$ results in a time-varying driving force $F = F_d \cos(\omega_d t)$.

The vibration amplitude at the drive frequency, assuming a steady-state response of the form $q(t) = A\cos(\omega_d t)$, is obtained by solving the cubic equation $A^2 = (F_d/(2m\omega_0))^2/[(\delta\omega - 3\alpha/(8\omega_0)A^2)^2 +$ $(\Gamma + (\eta/8)A^2)^2]$, with $\delta\omega = \omega_d - \omega_0$ being the frequency detuning. This expression captures the nonlinear resonator's amplitude-dependent response, where the resonance frequency shifts with increasing amplitude. When this equation yields three positive solutions for $q$, only the lowest and highest amplitude solutions are dynamically stable.

Figure 1 shows a representative measurement of $q$ versus $\delta f = \delta\omega/(2\pi)$ for the fundamental out-of-plane mode, driven at an optical power of $P = 22$ µW. Within the 130-Hz wide bistable region (shaded area), the steady-state amplitude depends on the initial conditions. In frequency-sweep experiments, the system exhibits hysteresis: Sweeping from low to high (high to low) frequencies drives the resonator into the high-amplitude (low-amplitude) solution. Our measurements match the theoretical solutions from the cubic amplitude equation (solid gray line), and follow the Duffing backbone (dashed gray line). Notably, the maximum amplitude difference between the device's two stable states in the bistable region span nearly three orders of magnitude, from 6 nm to 2,855 nm.

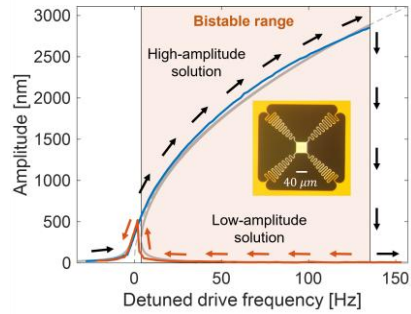

**Fig. 1: Bistability arising from Duffing nonlinearity.** Measured amplitude response of the silicon nitride trampoline resonator as a function of drive detuning, $\delta f$. A 130 Hz wide bistable region (shaded) is observed between two bifurcation points, where high- and low-amplitude solutions coexist due to the positive Duffing nonlinearity. Arrows indicate the direction of frequency sweeps. The experimentally measured responses for upward (low-to-high) and downward (high-to-low) frequency sweeps are shown in blue and orange, respectively. The solid gray line represents the theoretical solutions of the amplitude cubic equation described in the main text, while the dashed gray line depicts the Duffing backbone curve. Inset: optical image of the 50 nm-thick resonator, with a central membrane suspended by compliant serpentine springs.

Next, we consider the role of thermomechanical noise $\xi(t)$ in shaping the oscillator's response by extending the driving term to $F = F_d \cos(\omega_d t) + \xi(t)$. The effect of noise is most clearly observed in the power spectral density (PSD) or emission spectrum of the oscillator's motion. When the weakly damped oscillator is driven within the bistable regime, the PSD in the high-amplitude state exhibits two, symmetrically detuned sidebands[2]. These features are observed experimentally for a representative case of $\delta f = 27$ Hz, where the sidebands appear at $f_{SB} = 4.8$ Hz (Fig. 2a). To better understand their emergence, we move to a rotating reference frame at the drive frequency using the transformation $q(t) = X(t)\cos(\omega_d t) + Y(t)\sin(\omega_d t)$, where the quadratures $X(t)$ and $Y(t)$ evolve slowly on a characteristic timescale $\propto \Gamma^{-1}$ [2,48]. Substituting this expression into Eq. (1) and applying the rotating wave approximation ($\dot{X} \ll \omega_d X$, $\dot{Y} \ll \omega_d Y$), neglecting the noise term for now, we obtain the EOMs for the quadratures:

$$\dot{X} = \frac{\partial H^{\text{rw}}(X,Y)}{\partial Y} - \Gamma X$$

$$\dot{Y} = -\frac{\partial H^{\text{rw}}(X,Y)}{\partial X} - \Gamma Y$$

(2)

With the quadratures' energy landscape described by the rotating wave Hamiltonian:

$$H^{\text{rw}}(X,Y) = \frac{3\alpha}{32\omega_d}(X^2+Y^2)^2 - \frac{\delta\omega}{2}(X^2+Y^2) - \frac{F_d}{2m\omega_d}X$$

(3)

Figure 2b presents the vector flow (gray arrows) of the quadratures according to Eq. (2), overlaid with a contour plot of $H^{\text{rw}}(X,Y)$, calculated within the bistable regime. This visualization illustrates how any given initial condition evolves within the energy landscape, with those that lead to the low- or high-amplitude solutions defining their respective basins of attraction, separated by a separatrix (black line).

To gain deeper insight into the noise-induced dynamics within the high-amplitude basin, we project the measured oscillator trajectories onto the rotating-frame phase space (Fig. 2b, red curve). The trajectory exhibits stochastic, periodic motion around a fixed point. A zoomed-in view with color-coded time evolution is shown in Fig. 2c. This motion occurs on a time scale of $f_{\text{SB}}^{-1}$, illustrating how the sidebands originate from thermal fluctuations around stable points in phase space. The characteristics oscillation frequency can also be obtained by linearizing the quadratures EOMs[2], yielding $\omega_{\text{SB}} = \Gamma\sqrt{(3|y_0|^2 - \Delta\omega)(|y_0|^2 - \Delta\omega)}$ with normalized amplitude $|y_0| = q\sqrt{3\alpha(8\Gamma\omega_{\text{d}})^{-1}}$ and normalized detuning $\Delta\omega = \delta\omega\Gamma^{-1}$. The analytical sideband frequencies match the measured PSD (Fig. 2a, vertical dashed lines). Additionally, the unequal fluctuation extent along both quadrature axes indicates noise squeezing, manifested as an elongated trajectory in the phase portraits.

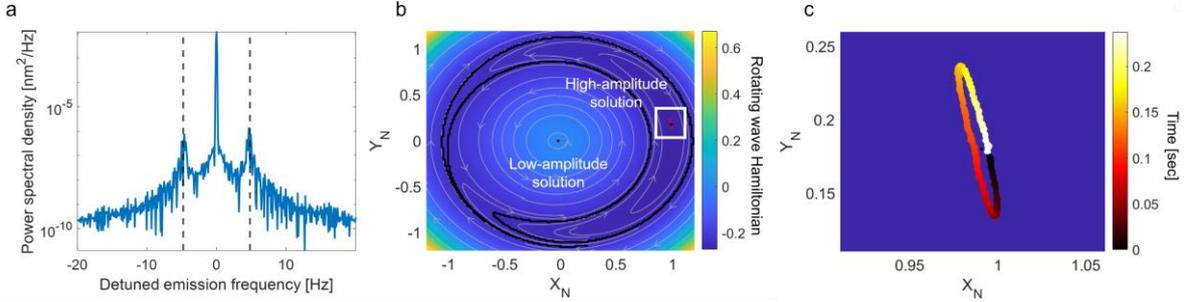

**Fig. 2: Noise-induced sidebands, noise squeezing, and trajectories in the rotating frame.** The resonator is driven within the bistable range at a detuning of 27 Hz from the resonance frequency. **a**, Power spectral density of the oscillations, showing a central peak at the drive frequency and two sidebands at ±4.8 Hz detuning. The theoretically predicted sideband frequencies are indicated by vertical dashed lines. **b**, Rotating-frame phase space representation of the oscillator. The color-coded background represents the normalized rotating-wave Hamiltonian, $H_N^{rw} = 3\alpha/(8\omega_d\delta\omega^2)H^{rw}$, plotted as a function of scaled quadratures $X_N$ and $Y_N$, which are obtained by normalizing the original quadratures $X$ and $Y$ by $\sqrt{8\omega_d\delta\omega/(3\alpha)}$. Gray arrows depict the vector field of the quadrature equations, and black points indicate the stable high- and low-amplitude solutions. The separatrix, which divides the two basins of attraction, is shown as a black line. The experimental trajectory of the resonator is plotted in red. **c**, Zoomed-in view of the trajectory, with time evolution being color-coded. The trajectory exhibits slow oscillatory motion around the stable solution, with a cycle of approximately 0.2 s, corresponding to the sideband frequency in **a**. The oblong shape of the trajectory reveals noise squeezing, where thermomechanical noise couples more strongly to one quadrature than the other.

Having established the dynamical picture in the rotating frame, we now examine manipulation of the oscillator state using an additional, weaker excitation. Driving oscillators with multiple excitations has been used to probe and engineer intermodal interactions, either by tuning the probe frequency to match the detuning between coupled modes or by directly exciting a second mode to enhance energy exchange[15,39,40,48,49]. Here, the total force $F = F_{\text{d}}\cos(\omega_{\text{d}}t) + F_{\text{p}}\cos(\omega_{\text{p}}t) + \xi(t)$ consists of the driving pump $F_{\text{d}}$, probe $F_{\text{p}}$ with frequency $f_{\text{p}} = \omega_{\text{p}}/2\pi$, and thermal noise $\xi(t)$. To complement our experimental observations and capture the oscillator response under varying drive conditions in the presence of thermomechanical noise, we numerically solve the stochastic differential EOM Eq. (1) using experimentally extracted parameters (see Methods).

As in Fig. 2, all studies presented in Fig. 3 are performed with the oscillator initialized in the high-amplitude state using $P = 22\ \mu\text{W}$ and $\delta f = 27$ Hz. We first characterize the oscillator response to a weak probe excitation scanned across frequencies near the primary drive. Figure 3a shows the simulated differential amplitude response, $\Delta A(f_{\text{p}}) = A^{(\text{probe on})}(f_{\text{p}}) - A^{(\text{probe off})}(f_{\text{p}})$, for a probe amplitude of $F_{\text{p}} = 0.08 F_{\text{d}}$, where $A^{(\text{probe off})}(f_{\text{p}})$ denotes the oscillator amplitude at $f_{\text{p}}$ in the absence of the probe excitation. The response exhibits clear resonant enhancement when the probe is applied near the sideband frequencies, where beating between the probe and the pump directly drives the sidebands in the rotating frame, as reflected in the quadrature equations including the second excitation (see Methods). This establishes that the sidebands provide access to probing and manipulating the resonator.

We next investigate the oscillator's response to the second excitation in the rotating frame, by analyzing the corresponding phase-space trajectories. When experimentally applying an acoustic probe near the right sideband (4.3 Hz detuning) at $F_p = 0.08 F_d$, the oscillator follows an extended trajectory along the high-amplitude basin boundary (Fig. 3b). Its measured PSD reveals resonant excitation of the sidebands and their harmonics (Fig. 3c). When the probe is increased to $F_p = 0.3 F_d$, the simulated trajectory in Fig. 3d shows the oscillator escaping the high-amplitude basin, with the probe overcoming the separatrix and pushing the system into the low-amplitude state.

To further investigate how the second excitation influences these transitions, we simulate the transition probability from the high- to the low-amplitude attractor as a function of the force ratio $F_p/F_d$ and frequency detuning $f_p - f_d$, obtaining the instability diagram shown in Fig. 3e. Thermomechanical noise leads to statistical behavior and thus probabilistic release rates, particularly near the transition zone boundaries. The transition rate increases with stronger perturbation and exhibits clear resonant enhancement when $|f_p - f_d| \approx 2 f_{SB}/n$, with $n \in \mathbb{Z}^+$. According to the rotating-frame nonlinear equations of motion (see Methods), pump-probe beating, manifested in the terms $\frac{F_p}{2m\omega_d}\sin(\omega_p - \omega_d)$ and $\frac{F_p}{2m\omega_d}\cos(\omega_p - \omega_d)$, induces amplitude modulation of the oscillator. Due to the intrinsic Duffing nonlinearity, this amplitude modulation leads to a temporally periodic modulation of the effective stiffness, enabling parametric-like excitation near sideband resonances. Therefore, the system exhibits lobes in the transition probability reminiscent of Arnold tongues[46,10].

Experimental histograms of the oscillator release rate as a function of the probe detuning $f_p - f_d$ for four different probe amplitudes (Fig. 3f) are consistent with the simulations in Fig. 3e, showing enhanced transitions at the sidebands ($n = 2$) for weak probes, and additional peaks at their second-harmonic ($n = 1$) and fractional ($n = 3$) resonances for stronger driving.

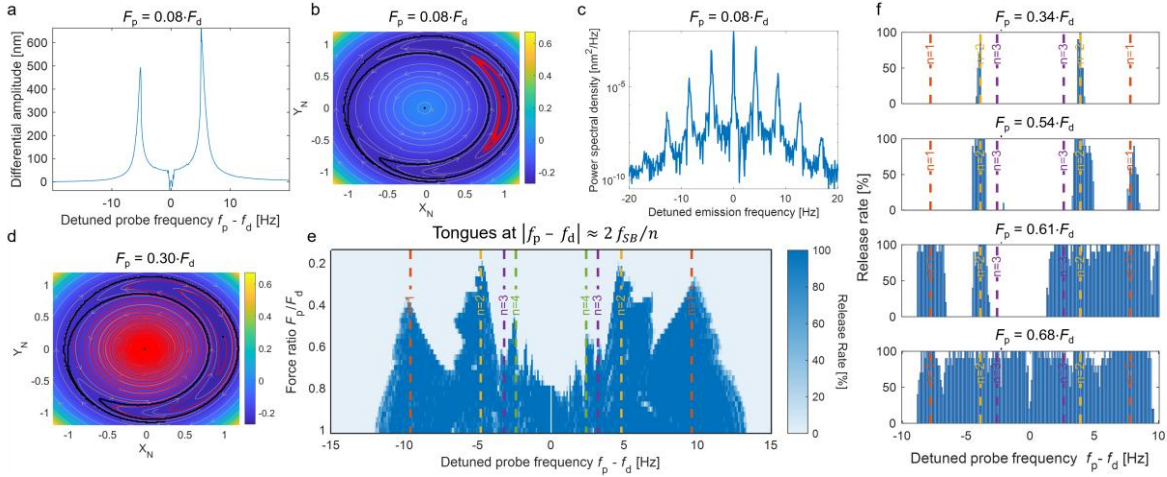

**Fig. 1: Manipulating resonator state in the rotating frame via a probe excitation**. **a**, Differential amplitude response from simulations, $\Delta A(f_p) = A^{(probe\,on)}(f_p) - A^{(probe\,off)}(f_p)$, showing the effect of adding a probe excitation to the primary drive. The pump corresponds to a laser power of 22 μW, applied at a detuning of $\Delta f = 27\,Hz$. The probe amplitude is 8% of the pump. The oscillator exhibits resonant enhancement when the probe is applied near sideband frequencies. **b**, Rotating-frame phase-space experimental trajectory with the probe applied at a detuning of 4.3 Hz (right sideband), at 8% of the pump amplitude. The resonator remains on the high-amplitude attractor, exhibiting elongated trajectories with slow amplitude modulation. **c**, Power spectral density of experimental trajectory in **b**, showing resonant excitation of sidebands and their harmonics. **d**, Rotating-frame phase-space trajectory (simulated) with the probe applied at a detuning of 4.8 Hz (right sideband), at 30% of the pump amplitude. The oscillator crosses the separatrix, driving the system into the low-amplitude state. **e**, Simulated instability diagram: transition rate from the high- to low-amplitude attractor as a function of force ratio $F_p/F_d$ and frequency detuning $f_p - f_d$. The transition is probabilistic due to thermomechanical noise. The release rate increases with stronger perturbation or when $|f_p - f_d| \approx 2 f_{SB}/n$, where $n$ is an integer, demonstrating nonlinear internal route to parametric-like behavior. **f**, Experimental histograms of the release rate versus detuning $f_p - f_d$ for four different $F_p$ values. The results align with **e**, with enhanced transitions at the sideband (*n* = 2)

for weak perturbations and additional peaks at the second harmonic (*n* = 1) and fractional (*n* ≥ 3) resonances for stronger driving. In **e** and **f**, the probe was applied for 3 s, followed by a pause of 5 s allowing the resonator to ring down if released, before measuring the amplitude to determine whether a transition to the low-amplitude state occurred.

So far, we have considered the nonlinear dynamics of a single vibrational mode. However, our micromechanical resonator supports many vibrational modes, each of which can be effectively described by the same general framework of a driven damped Duffing resonator. We extend our analysis to a system comprising two such nonlinear modes at different frequencies, coupled through displacement-induced tension. Specifically, we consider a higher-order flexural mode at $f_{0,3} = 17{,}958\,\text{Hz}$ in addition to the fundamental mode at $f_{0,1} = 5{,}862\,\text{Hz}$, with their simulated mode shapes shown in Fig. 4a. These modes exhibit three or one antinodes along the spring's diagonal, with the central membrane located at an antinode, enabling strong spatial overlap and efficient optical drive coupling. Mode 3 also exhibits bistability (Fig. 4b) where a low-to-high frequency sweep ($P = 53\,\mu W$) reveals a 2.7 Hz-wide bistable region (shaded). Therefore, the resulting dynamics are well captured by two stochastic, nonlinearly coupled Duffing EOMs with external drives $F_i(t)$ and thermomechanical noise $\xi_i(t)$:

$$\ddot{q}_1 + 2\Gamma_1 \dot{q}_1 + \omega_{0,1}^2 q_1 + \alpha_1 q_1^3 + \eta_1 \dot{q}_1 q_1^2 + \kappa_{13} q_1 q_3^2 = m_1^{-1}(F_1(t) + \xi_1(t)) \quad (4a)$$

$$\ddot{q}_3 + 2\Gamma_3 \dot{q}_3 + \omega_{0,3}^2 q_3 + \alpha_3 q_3^3 + \eta_3 \dot{q}_3 q_3^2 + \kappa_{31} q_3 q_1^2 = m_3^{-1}(F_3(t) + \xi_3(t)) \quad (4b)$$

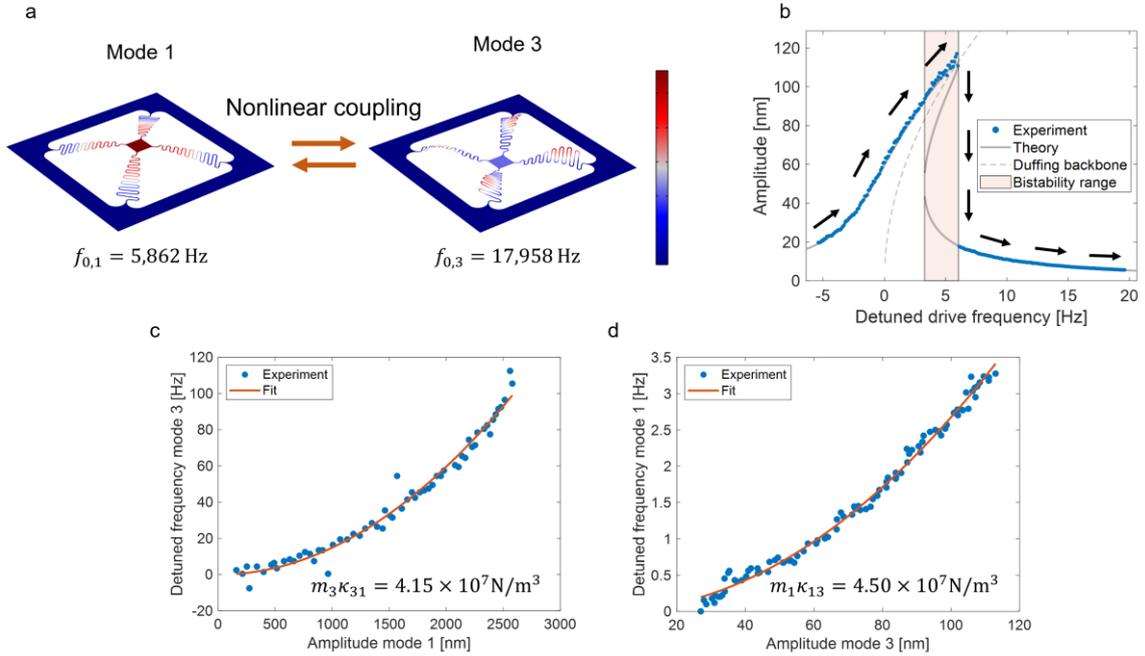

**Fig. 4: Nonlinear coupling between flexural modes via displacement-induced tension**. **a**. Simulated out-of-plane flexural mode shapes (COMSOL Multiphysics) of the fundamental mode (mode 1, left) at $f_{0,1} = 5{,}862\,Hz$ and the third-order mode (mode 3, right) at $f_{0,3} = 17{,}958\,Hz$. Colors indicate the magnitude of out-of-plane displacement. The modes exhibit 1 and 3 antinodes along the spring's diagonal, respectively, with the central membrane located at an antinode, enabling strong spatial overlap and efficient optical drive coupling. **b**. Amplitude response of mode 3 under optical drive ($P = 53\,\mu W$) during a low-to-high frequency sweep, revealing a 2.7 Hz-wide bistable region (shaded) due to Duffing nonlinearity. The experimental fit (solid gray) and associated backbone curve (dashed gray) are overlaid. **c**. Frequency shift of mode 3 when mode 1 is driven, demonstrating nonlinear dispersive coupling. **d**. Frequency shift of mode 1 when mode 3 is driven. The observed frequency shifts follow $\omega_{0,\text{shifted},i}^2 = \omega_{0,i}^2 + (3/4)\alpha_i A_i^2 + (1/2)\kappa_{ij} A_j^2$, from which the coupling coefficients shown in the panels are extracted.

The natural resonance frequency $\omega_{0,i}$ of mode $i$ is shifted to an effective frequency $\omega_{0,\text{shifted},i}$ due to the self- ($\alpha_i$) and cross-nonlinearity ($\kappa_{ij}$). Assuming harmonic motion, the frequency shift depends on the squared modal amplitudes $A_{i,j}^2$ (see Supplementary Note 3), leading to:

$$\omega_{0,\text{shifted},i}^2 = \omega_{0,i}^2 + \frac{3}{4}\alpha_i A_i^2 + \frac{1}{2}\kappa_{ij} A_j^2 \quad (5)$$

Therefore, by sweeping an excitation across the resonance of one mode only and recording its displacement, we extract the nonlinear cross-coupling coefficients $m_1 \cdot \kappa_{13} = 4.50 \times 10^7 \text{N/m}^3$ and $m_3 \cdot \kappa_{31} = 4.15 \times 10^7$ N/m$^3$ based on the associated frequency shift of the other mode (Fig 4c,d).

We now examine how intermodal coupling can be leveraged to control escape dynamics through intermodal interactions. After preparing the fundamental mode again in its high-amplitude state, choosing a larger detuning (120 Hz), we sweep a second excitation across mode 3, producing probe-frequency dependent PSDs shown in Fig. 5a. The effective probe force scaled to 40% of the fundamental mode's drive (defined according to $F_i/m_i$), reveals rich nonlinear dynamics.

Strong excitation of mode 1 shifts mode 3 by ~130 Hz via nonlinear coupling (Eq. (5)). When mode 3 is only thermally excited due to a far-detuned probe, the fundamental mode amplitude remains constant at around ~1,400 nm (Fig. 5b). Nonlinearities of the device produce spectral wave-mixing signatures, including a notable six-wave mixing peak at $4f_d - f_p$. As the probe frequency approaches shifted mode 3, its amplitude increases, while mode 1's amplitude decreases (Fig. 5b). Simultaneously, the fundamental mode's sideband frequencies change from 4.5 Hz to 7.5 Hz (Fig. 5c, vertical lines). Interpreting the sidebands as the detuning between the drive frequency and the nonlinearly-shifted, thermally-excited resonance of mode 1, this observation can be explained by Eq. (5): Exciting mode 3 blueshifts mode 1's effective resonance via intermodal coupling, which increases the sideband detuning and reduces mode 1's amplitude response (Fig. 5b, inset).

When the probe exceeds the shifted mode 3 by approximately $f_p - f_{0,\text{eff},3} \approx 4.5$ Hz, the fundamental mode transitions from the high-amplitude to the low-amplitude state. This can be seen by an abrupt change in resonance frequency (vertical arrows in Fig. 5a) and amplitude (color in Fig. 5a) of both modes. To better understand the mechanism for the intermodal escape dynamics, we assume mode 3's displacement to contain two frequency components:

$$q_3(t) = A_{0,\text{eff},3} \cos(\omega_{0,\text{eff},3} t) + A_p \cos(\omega_p t), \tag{6}$$

where $A_{0,\text{eff},3}$ is the amplitude at the shifted, noise-excited resonance, and $A_p$ is the amplitude at the probe. This two-frequency excitation induces a time-dependent modulation of the resonance frequency of mode 1 via nonlinear coupling (Supplementary Note 4). Defining the shifted resonance frequency as in Eq. (5), mode 1's time-modulated resonance frequency becomes:

$$\omega_{0,\text{eff},1}^2(t) = \omega_{0,\text{shifted},1}^2 [1 + \delta \cos(\delta \omega_3 t)], \tag{7}$$

where $\delta \omega_3 = \omega_p - \omega_{0,\text{eff},3}$ is the detuning between the probe frequency and mode 3's shifted resonance, and $\delta$ is the modulation depth given by:

$$\delta = \frac{\kappa_{13} A_{0,\text{eff},3} A_p}{\omega_{0,\text{shifted},1}^2}. \tag{8}$$

Therefore, nonlinear cross-mode coupling can cause parametric modulation of the fundamental mode's resonance frequency due to beating between frequency components of the higher-order mode.

To quantify this parametric modulation in our experiment, we evaluate the modulation depth according to Eq. (8) as a function of detuning from the probe $\omega_p/2\pi$, effectively sweeping $\omega_{0,\text{eff},3}$. As the system approaches the release point, $\delta$ increases (Fig. 5d), with the strongest modulation occurring at low frequencies, consistent with the range identified in Fig. 3e as relevant for inducing escape. We also observe enhanced modulation at the shifting sidebands of the fundamental mode.

However, unlike previously, the fundamental mode's sidebands are not directly driven. To isolate the effect of parametric modulation near the sidebands without directly exciting them, we simulate its response using Eq. (1) with a time-varying resonance frequency according to Eq. (7) and assuming $\delta = 2.5 \times 10^{-4}$ (within the relevant experimental range according to Fig. 5d). For every perturbation frequency, the oscillator is initialized in the high-amplitude state. After three-second-long parametric modulation of the resonance frequency (allowing ringdown of the oscillator if released), the PSD is then computed, producing the emission spectra in Fig. 5e. Spectral features correspond to noise-actuated sidebands (horizontal), along with peaks at the modulation frequency, its harmonics, and mixing products with the drive (diagonal). Escape occurs when the parametric modulation approaches the sideband range (blue vertical bands).

Increasing the modulation depth broadens the release range beyond the sidebands into a continuum, consistent with the behavior shown in Fig. 3e. This interpretation together with the chosen probe magnitude explains the experimental release at $f_p - f_{0,\text{shifted},3} \approx 4.5$ Hz rather than the shifted sideband. Thus, we establish that spectral beating in a higher-order mode can induce parametric modulation of the fundamental mode, and that such modulation, when resonant with the fundamental mode's sidebands, can trigger escape from the high-amplitude state.

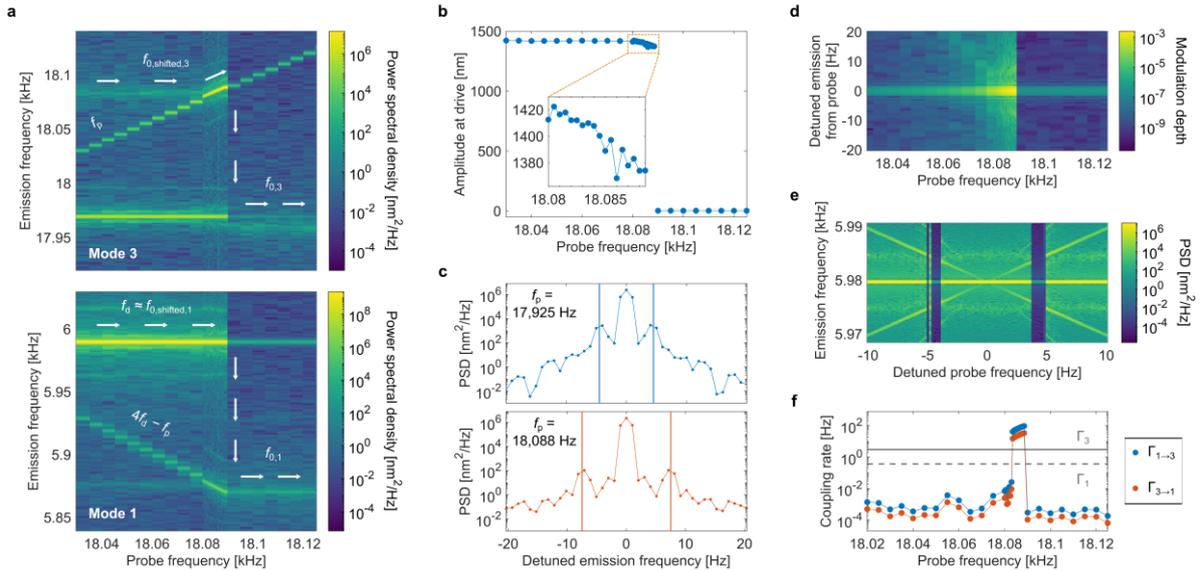

**Fig: 5 Control of escape dynamics via intermodal coupling and parametric modulation. a**, Probe-frequency-dependent power spectral density of mode 3 (top) and mode 1 (bottom) as the probe is swept across the resonance of mode 3 after preparing mode 1 in the high-amplitude state by the drive. As the probe resonantly enhances mode 3's amplitude, escape from the high-amplitude state of mode 1 occurs, marked by abrupt shifts in the modes' frequencies (arrows) and amplitudes (color-coded). **b**, Displacement amplitude of mode 1 versus probe frequency. The inset highlights a gradual decrease in amplitude preceding the escape. **c,** Power spectral densities of mode 1 showing sideband evolution: from 4.5 Hz when the probe is far detuned (top) to 7.5 Hz near the escape threshold (bottom), consistent with amplitude-dependent softening. **d**, Modulation depth of mode 1's resonance frequency calculated from experimental values due to parametric modulation by the beating components of mode 3, following Eq. (8). The modulation depth increases as the system approaches escape, particularly in the low-frequency range. **e**, Simulated PSDs from a single-mode model with time-varying resonance frequency. Features appear at the modulation frequency, its harmonics, and sidebands. Escape is indicated by the disappearance of spectral power across the full range of emission frequencies shown, near sideband frequencies (vertical strips). **f**, Calculated intermodal coupling strength versus probe frequency, showing instantaneous energy exchange rates exceeding the dissipation rates by one order of magnitude before the escape point, with a five-order-of-magnitude drop upon release.

An intriguing aspect of the abrupt amplitude drop upon release from the high- to low-amplitude states is the associated change in the coupling terms. Unlike linear coupling with amplitude-independent energy exchange rates, nonlinear coupling depends strongly on the mode amplitudes, thereby affecting the temporal dynamics and altering energy dissipation in these systems[13,36,50] (see Supplementary Note 5). We define an effective coupling strength from the energy exchange rate's dominant spectral component to capture the key behavior. As shown in Fig. 5f, this coupling strength exceeds both modes' dissipation rates by one order of magnitude before release, then drops by over five orders of magnitude afterwards. This large variation demonstrates dynamical tuning of intrinsic nonlinear coupling between modes.

## Conclusion

We demonstrate how the combination of Duffing bistability, intermodal coupling, and thermomechanical noise reveal a new class of optomechanical nonlinearities in micromechanical resonators. Within the bistable regime, we show that noise-induced sidebands, typically considered passive signatures of thermal motion, can be actively harnessed to induce transitions between coexisting vibrational states using a weak secondary excitation. These inherently stochastic transitions reflect the joint influence of coherent modulation and thermal fluctuations.

Extending this mechanism to multimode interactions, we find that nonlinear coupling between vibrational modes can similarly mediate transitions, through parametric modulation driven by spectral beating in a higher-order mode. When this modulation resonates with the sidebands of the fundamental mode, it triggers escape from the high-amplitude state without direct sideband driving. This process is accompanied by a change in intermodal coupling strength of five orders of magnitude, highlighting how in nonlinear systems, the coupling itself can become a dynamically tunable quantity.

Together, these findings point to new opportunities for designing reconfigurable optomechanical networks, where energy transport, resonance conditions, and coupling strength can be modulated on demand. The ability to harness stochastic fluctuations for controlled switching also suggests future applications in nonreciprocal signal routing, noise-enhanced sensing, and dynamical logic operations. More broadly, our results highlight thermally-driven nonlinear dynamics as a powerful resource for precision control in mesoscopic systems.

## Methods

### Device fabrication

The spring-supported trampoline resonator is patterned into a low-stress silicon nitride membrane (Norcada, NXA10040C, 100 nm thickness) via electron-beam lithography (ZEP 520A, D.R. 1.5, aquaSAVE, development in ZED-N50) and transferred via inductively coupled reactive-ion etching (pseudo-Bosch etch using $SF_6/C_4F_8$ chemistry). Due to the fragility of the etched device, the resist is stripped in a weak $O_2$ plasma for 90 minutes. The final thickness of the membrane is 50 nm given the etching conditions.

### Experimental setup

All measurements are conducted using an automated experimental setup. Specifically, the mixed-domain oscilloscope (Tektronix, MDO32) and the arbitrary waveform generators (Keysight, 33220A AWG & Siglent, SDG 2042x) are controlled with dedicated MATLAB scripts.

### Detection of displacement with common-path interferometry

To detect lightsail displacement, free-space common path interferometer is employed. To generate the sample and reference probe beams, light from a stabilized HeNe (Thorlabs HRS015B, 632.8 nm, polarized, 1.2 mW) is split using a binary phase grating DBS (Holor/Or, fused silica, anti-reflection coated) and coupled into the microscope (Zeiss, Axio Observer Z1). The laser attenuation is set using neutral density filters to probe the pad motion with a power of ~1 µW. The reflected HeNe beams are detected by a temperature-compensated silicon avalanche photodiode (Thorlabs APD410A) and recorded by the mixed-domain oscilloscope. Laser power was measured by placing an optical power meter (Thorlabs, PM16-120) at the sample plane.

### Laser excitation path

The device is driven by a collimated continuous-wave (CW) argon-ion laser at 514 nm (Innova 70c). Intensity modulation is applied using an acousto-optic modulator (AOM) (NEOS 23080-1), connected to a feedback-controlled RF Driver (Isomet, 232A-2). The modulation feedback loop is implemented using a servo controller (Newport, MKS, LB1005), which monitors the power of the modulated beam with a silicon photodiode (ams-OSRAM, BPW 34 S-Z), soldered in parallel to a 1 kΩ resistor. The photodiode output is amplified by a transimpedance amplifier (TIA) (Edmunds Optics, SN 57-601). This signal is used to adjust the input to the AOM in real time, ensuring that the modulation matches the desired waveform set by the arbitrary waveform generator (AWG).

**Data analysis**

Driving the device on or near the fundamental resonance leads to large oscillation amplitudes, exceeding the several HeNe wavelengths (632.8 nm). This results in phase wrapping of the interferometric signal, which requires a sufficiently high sampling rate to avoid aliasing (10 MHz for data shown in Fig. 2, 1 MHz for data shown in Fig. 5). Simultaneously, due to memory constraints of the oscilloscope, this limits the duration of data acquisition to ten seconds for results shown in Fig. 2 and 3, and one second for results shown in Fig. 5, resulting in an emission frequency resolution of 0.1 Hz and 1 Hz, respectively. We record the full-time domain interferometric signal and demodulate it post-acquisition after identifying folding points (rather than the true extrema of oscillation), normalizing them to $\pm 1$, applying an $\cos^{-1}(\cdot)$ operation to the normalized signal and converting it to units of displacement.

To obtain the power spectral densities shown in Fig. 2, Fig. 3 and Fig. 5, we use the pwelch command in MATLAB, applying the Hann window to the data. We do not spectrally average the data in order to maintain the frequency resolution set by the data acquisition duration.

**Drive and probe force**

The radiation pressure force exerted on the resonator is given by $F = Pc^{-1}(2R + A)$, where $P$ is the optical power, $c$ is the speed of light, $R$ is the reflectance, and $A$ is the absorption coefficient. This expression accounts for both the reflected and absorbed components of momentum transfer. For our device, driven at a wavelength of 514 nm, we have experimentally determined $R = 0.4$ and $A \approx 10^{-4}$ (Ref. [51]). Given the negligible magnitude of $A$ compared to $R$, absorption can be neglected in our analysis. Moreover, the low absorption ensures that photothermal forces are insignificant (see Ref. 49), validating our use of a purely radiation-pressure-driven model in the main text.

In addition, we employ an bone conducting transducer (Dayton Audio BCT-3) with a Tungsten cube (1.5" x 1.5" x 1.5", 1 kg) on top, placed next to the vacuum chamber to probe the driven modes in our experiments. Specifically, while the drive in all experiments comes from optical forces, the probe is applied mechanically using a second function generator (Siglent SDG 2042x). To calibrate the voltage-to-force conversion of the piezoelectric actuator, we compared its effect on the device to that of the independently calibrated optical force.

**Full equations of motion in the rotating frame**

The EOMs of the quadratures $X(t)$ and $Y(t)$ in the rotating frame, including thermomechanical noise $\xi(t)$ and a probe with amplitude $F_p$ and angular frequency $\omega_p$, are given by:

$$\dot{X}(t) = -\Gamma X(t) - Y(t)\left((\omega_d - \omega_0) - \frac{3\alpha}{8\omega_d}(X^2(t) + Y^2(t))\right) + \frac{F_p}{2m\omega_d}\sin\left((\omega_p - \omega_d)t\right) + \Im\{\bar{\xi}_y\} \quad (1)$$

$$\dot{Y}(t) = -\Gamma Y(t) + X(t)\left((\omega_d - \omega_0) - \frac{3\alpha}{8\omega_d}(X^2(t) + Y^2(t))\right)$$

$$+ \frac{F_d}{2m\omega_d}\left(1 + r\cos\left((\omega_p - \omega_d)t\right)\right) + \Re\{\bar{\xi}_y\}$$

The terms $\Im\{\bar{\xi}_y\}$ and $\Re\{\bar{\xi}_y\}$ denote the scaled imaginary and real parts, respectively, of the noise term in the rotating frame $\bar{\xi}_y = \frac{1}{m\omega_d}\xi(t)e^{i\omega_d t}$, and $r = F_p/F_d$ is the force ratio.

These equations reveal that the beating between the probe and pump induces a time-periodic modulation of the effective drive, which can resonantly excite the oscillator near sideband frequencies in the rotating frame. This modulation, in combination with thermomechanical noise, facilitates escape from the high-amplitude basin of attraction.

**Stochastic differential equation simulation**

To numerically solve the stochastic EOMs that include thermomechanical noise, we employ a Runge–Kutta method of strong order 1[52], suitable for stochastic differential equations (SDEs) in the Itô form:

$$dX = a(t, X)\, dt + b(t, X)\, dW \tag{5}$$

where $X(t) \in \mathbb{R}^n$ is the state vector, $a(t, X)$ is the deterministic drift term, $b(t, X)$ is the diffusion coefficient, and $dW$ is the increment of a Wiener process.

Given a timestep $\Delta t$ and the state $X_k = X(t_k)$, the value at the next step, $X_{k+1}$, is computed as follows:

$$K_1 = a(t_k, X_k)\,\Delta t + \left(\Delta W_k - S_k\sqrt{\Delta t}\right) \cdot b(t_k, X_k),$$

$$K_2 = a(t_{k+1}, X_k + K_1)\,\Delta t + \left(\Delta W_k + S_k\sqrt{\Delta t}\right) \cdot b(t_{k+1}, X_k + K_1), \tag{5}$$

$$X_{k+1} = X_k + \frac{1}{2}(K1 + K2),$$

Where $\Delta W_k \sim N(0, \Delta t)$ is a normally distributed random variable representing the Wiener increment, and $S_k = \pm 1$ with equal probability 1/2.

This method was chosen for its balance of speed, accuracy, and simplicity, making it well-suited for the large-scale simulations in this work[53].

**Data availability**

The data supporting the findings of this study are available from the corresponding author upon request.

## Acknowledgements

This work was supported by the Air Force Office of Scientific Research under grant FA2386-18-1-4095 and the Breakthrough Starshot Initiative.

LM acknowledges support of the Fulbright Israel Postdoctoral Fellowship and thanks Yoni Schattner for enjoyable and valuable discussions. The authors acknowledge helpful discussions with Keith Schwab.

## Author information

Authors and Affiliations

Department of Applied Physics and Materials Science, California Institute of Technology, Pasadena, CA 91125, USA

Lior Michaeli, Ramon Gao, Michael D. Kelzenberg, Claudio U. Hail, Adrien Merkt, John E. Sader

 & Harry A. Atwater

Contributions

L.M. conceived the ideas and developed the theory with insights from R.G. and H.A.A. L.M. performed the numerical simulations. R.G. designed and fabricated the device. L.M. and R.G. built the setup, with inputs from M.D.K. R.G. devised and conducted the measurements, with inputs from L.M. R.G. and L.M. performed the data analysis and interpretation. C.U.H. and J.E.S. contributed to discussions. J.E.S. provided insights into the theory. L.M. primarily wrote the manuscript, with R.G. contributing to writing and revision of the manuscript, and H.A.A. providing valuable feedback. H.A.A. supervised the project.

Corresponding author

Correspondence to Harry A. Atwater.

## Ethics declarations

Competing interests

The authors declare no competing interests.

## Supplementary information

File is attached.

Supplementary Notes 1-5, Supplementary Figures 1-2 and Supplementary References.

# Supplementary Information: Optically Actuated Transitions in Multimodal, Bistable Micromechanical Oscillators

Lior Michaeli[†], Ramon Gao[†], Michael D. Kelzenberg, Claudio U. Hail, John E. Sader, and Harry A. Atwater[*]

Department of Applied Physics and Materials Science, California Institute of Technology, Pasadena, CA 91125, USA

[†]These authors contributed equally: Lior Michaeli, Ramon Gao

[*] *haa@caltech.edu*

# Supplementary Note 1: Rotating frame analysis for a two-tone driven Duffing oscillator

We consider the nonlinear dynamics of a Duffing oscillator with displacement coordinate $q(t)$, subjected to two external excitations: a primary excitation denoted as drive with amplitude $F_d$ and frequency $\omega_d$, and a secondary excitation denoted as probe with amplitude $F_p$ and frequency $\omega_p$. The equation of motion also includes additive thermal noise:

$$\ddot{q} + 2\Gamma\dot{q} + [\omega_0^2 + \alpha q^2]q = m^{-1}\left(F_d\cos(\omega_d t) + F_p\cos(\omega_p t) + \xi(t)\right) \tag{1}$$

Where $\omega_0$ is the natural frequency of the oscillator, $\Gamma$ is the linear damping rate, $\alpha$ is the nonlinear stiffness coefficient, and $m$ is the effective mass of the oscillator.

The stochastic force $\xi(t)$ accounts for thermal fluctuations and satisfies the fluctuation-dissipation theorem. It is a zero mean Gaussian process, $\langle\xi(t)\rangle = 0$, with autocorrelation function given by $\langle\xi(t)\xi(t')\rangle = 4\Gamma k_B T m \delta(t-t')$, where $k_B T$ is the thermal energy.

To capture the slow envelope dynamics, we move to a rotating frame at the primary drive frequency $\omega_d$, and define the complex amplitude $y(t)$:

$$q(t) = \sqrt{\frac{2\omega_d \Gamma}{3\alpha}}[y(t)e^{i\omega_d t} + y^*(t)e^{-i\omega_d t}] \tag{2}$$

Assuming that the envelope varies slowly on timescale of $\omega_d^{-1}$, i.e., $\dot{y}(t) \ll \omega_d y(t)$, we apply the rotating wave approximation by substituting Eq. (2) into Eq. (1), and retaining only near-resonant terms. This yields the envelope equation:

$$\dot{y} = -\Gamma(1 + i\Delta\omega - i|y|^2)y - i\Gamma\beta\left(1 + re^{i(\omega_p - \omega_d)t}\right) - i\Gamma\xi_y(t) \tag{3}$$

Where we have defined:

$$\beta = \frac{F_d}{m}\sqrt{\frac{3\alpha}{32\omega_d^3\Gamma^3}} \ ; \ \delta\omega = \omega_d - \omega_0 \ ; \ \Delta\omega = \frac{\delta\omega}{\Gamma} ; \ \xi_y(t) = \frac{1}{m}\sqrt{\frac{3\alpha}{8\omega_d^3\Gamma^3}}\xi(t)e^{-i\omega_d t} \ ; \ r = \frac{F_p}{F_d} \tag{4}$$

**Steady-state amplitude in the single-tone limit**

In the absence of noise and the second excitation ($\xi(t) = 0, F_p = 0$), the system reaches a steady state $y(t) = y_0$. Equation (3) then becomes:

$$0 = -\Gamma(1 + i\Delta\Omega - i|y_0|^2)y_0 - i\Gamma\beta \tag{5}$$

Solving for $y_0$, we find:

$$y_0 = \frac{-i\beta}{1 + i\Delta\omega - i|y_0|^2} \tag{6}$$

Alternatively, one may express the steady-state solution directly in terms of the physical displacement amplitude $q$ by substituting back into the original lab-frame equation. Including also the effect of nonlinear damping $\eta q^2 \dot{q}$, we obtain the amplitude–frequency relation:

$$A^2 = \frac{\left(\frac{F_d}{2m\omega_0^2}\right)^2}{\left(\frac{\omega_d - \omega_0}{\omega_0} - \frac{3}{8}\frac{\alpha}{\omega_0^2}A^2\right)^2 + \left(\frac{\Gamma}{\omega_0} + \frac{1}{8}\frac{\eta}{\omega_0}A^2\right)^2} \tag{7}$$

Where $A$ denotes the steady-state amplitude of harmonic motion according to $q(t) = A\cos(\omega_d t)$. The associated phase shift $\phi$ between the drive and the response is given by:

$$\tan(\phi) = \frac{\Gamma + \frac{1}{8}\eta A^2}{\omega_d - \omega_0 - \frac{3}{8}\frac{\alpha}{\omega_0}A^2} \tag{8}$$

**Sideband response from linearized fluctuations**

To analyze small fluctuations around the steady-state, we write the displacement as $y(t) = y_0 + \delta y(t)$, where $\delta y(t)$ represent small perturbation around the steady-state amplitude $y_0$. Substituting it into Eq. (3) and linearizing in $\delta y$, we obtain:

$$\delta \dot{y} = -\Gamma(1 + i\Delta\omega - 2i|y_0|^2)\delta y - i\Gamma y_0^2 \delta y^* \tag{9}$$

This equation describes a pair of coupled first-order differential equations for $\delta y$ and $\delta y^*$, which can be solved to obtain the fluctuation spectrum. When written in matrix form, the eigenvalues of the resulting system determine the sideband frequencies:

$$\omega_{SB} = \Gamma\sqrt{(3|y_0|^2 - \Delta\omega)(|y_0|^2 - \Delta\omega) + 1} \tag{10}$$

With the amplitude $|y_0| = q\sqrt{3\alpha(8\Gamma\omega_d)^{-1}}$. This expression captures the oscillation frequency of small fluctuations about the driven steady-state, which manifest as sidebands relative to the drive frequency in the lab frame. We note that for our device and experiments carried out, $(3|y_0|^2 - \Delta\omega)(|y_0|^2 - \Delta\omega)$ is typically much larger than one, allowing us to neglect the latter.

**Real-valued quadrature formulation**

Whereas the derivation above used complex scaled amplitude in rotating wave, it is useful to derive the equation of motion for the real-valued quadratures $X(t)$ and $Y(t)$, which are conjugate variables, defined according to:

$$q(t) = X(t)\cos(\omega_d t) + Y(t)\sin(\omega_d t) \tag{11}$$

The quadratures relate to $y(t)$ via:

$$X(t) = \sqrt{\frac{8\omega_d \Gamma}{3\alpha}} \Re\{y(t)\}$$
$$Y(t) = -\sqrt{\frac{8\omega_d \Gamma}{3\alpha}} \Im\{y(t)\} \tag{12}$$

Expressing $y(t)$ in terms of $X(t)$ and $Y(t)$, we get:

$$y = \sqrt{\frac{3\alpha}{8\omega_d \Gamma}}(X(t) - iY(t)) \tag{13}$$

Substituting into Eq. (3), we obtain:

$$\dot{X}(t) = -\Gamma X(t) - Y(t)\left((\omega_d - \omega_0) - \frac{3\alpha}{8\omega_d}(X^2(t) + Y^2(t))\right) + \frac{F_p}{2m\omega_d}\sin\left((\omega_p - \omega_d)t\right)$$
$$+ \Im\{\bar{\xi}_y\}, \tag{14}$$

$$\dot{Y}(t) = -\Gamma Y(t) + X(t)\left((\omega_\text{d} - \omega_0) - \frac{3\alpha}{8\omega_\text{d}}(X^2(t) + Y^2(t))\right)$$
$$+ \frac{F_\text{d}}{2m\omega_\text{d}}\left(1 + r\cos\left((\omega_\text{p} - \omega_\text{d})t\right)\right) + \Re\{\bar{\xi}_y\},$$

with $\bar{\xi}_y = \frac{1}{m\omega_\text{d}}\xi(t)e^{i\omega_\text{d} t}$.

The quadrature equations reveal that the effective resonance frequency in the rotating frame shifts with amplitude due to the cubic Duffing nonlinearity. This is evident in the coefficients of $Y$ and $X$ in Eq. (14), where the stiffness depends on the amplitude. The positive sign of $\alpha$ in our system leads to resonance softening, consistent with the observed shift in sideband frequency from 4.8 Hz in the single-tone case (Fig. 2) to 3.9 Hz under strong two-tone excitation (Fig. 3e).

The addition of a second excitation (probe) introduces amplitude modulation via the $\cos\left((\omega_\text{p} - \omega_\text{d})t\right)$ and $\sin\left((\omega_\text{p} - \omega_\text{d})t\right)$ terms, producing beat-frequency oscillations. These modulate the oscillator's amplitude in time, and, through the nonlinear stiffness, induce a time-dependent restoring force. This dynamic modulation enables sideband excitation and gives rise to parametric-like behavior, as reflected by the appearance of Arnold tongue structures in both numerical simulations and experimental data (Fig. 3).

## Supplementary Note 2: Linear and nonlinear parameters of the vibrational modes

To extract the linear properties of the vibrational modes, we analyze the thermal noise fluctuations by recording the oscillator's time-domain signal and computing its power spectral density. The resonance peaks of the modes are fitted with a Lorentzian line shape, from which we extract the resonance frequency $f_{0,i}$, dissipation rate $\Gamma_i$, and stiffness $k_i$ (Fig. S1). The stiffness is determined from the area under the thermal noise spectrum, which gives the variance of the mechanical displacement $\langle q_i^2 \rangle$. Using the equipartition theorem, the stiffness is calculated as $k_i = k_\text{B}T/\langle q_i^2 \rangle$, where $k_\text{B}$ is Boltzmann's constant and T is temperature.

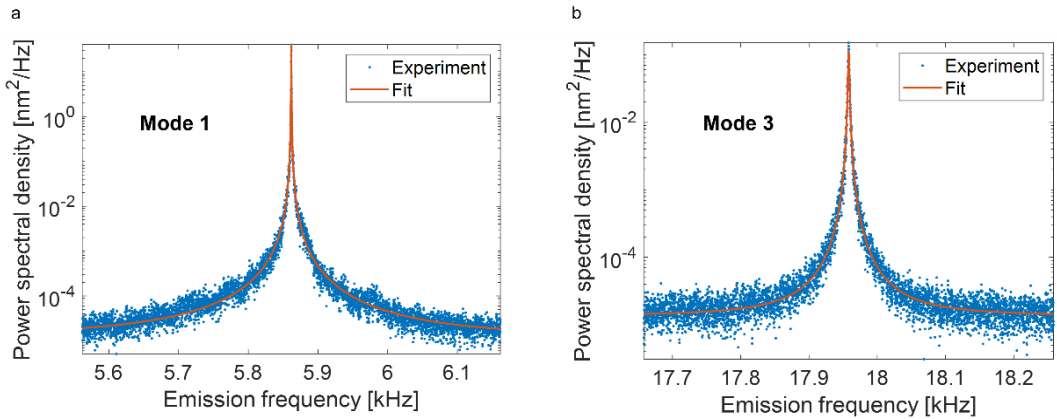

**Fig. S1: Thermal power spectral density (PSD) of the measured modes in the absence of external driving.** Panels (a) and (b) show the PSD of mode 1 and mode 3, respectively. The linear dynamical parameters - resonance frequency, damping rate, and stiffness - are extracted by fitting the thermal peak to a Lorentzian profile.

The nonlinear stiffness coefficient $\alpha$ is determined from fitting the experimental measurement of the shifted resonance frequency versus displacement amplitude for several magnitudes of optical forces (powers) to the Duffing backbone equation[1]:

$$\omega_{0,\max} = \omega_0 + \frac{3}{8}\frac{\alpha}{\omega_0}A^2. \tag{15}$$

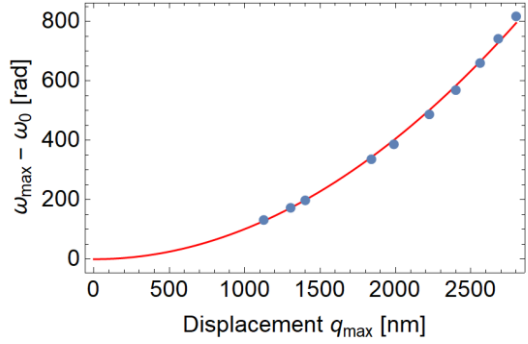

**Fig. S2: Extraction of nonlinear stiffness coefficient via fit to Duffing backbone equation.** Each data point represents a different optical driving force, which results in a specific pair of maximally shifted resonance frequency $\omega_{0,max}$ and corresponding displacement amplitude $A$.

The nonlinear damping coefficient is determined by the fit of the experimental Duffing curve to the theoretical Duffing curve in Fig. 1, given precise calibration of the optical force.

All extracted linear and nonlinear parameters for the two studied modes are summarized in Table 1.

| Mode | $f_0$ [Hz] | $k$ [N/m] | $m$ [kg] | $\alpha \cdot m$ [N/m³] | $\eta \cdot m$ [Ns/m³] | $\Gamma$ [Hz] | Q-factor | $\kappa_{ij} \cdot m_i$ [N/m³] |
|---|---|---|---|---|---|---|---|---|
| 1 | 5,862 | $2.48 \times 10^{-4}$ | $1.83 \times 10^{-13}$ | $1.83 \times 10^6$ | $2.13 \times 10^{-1}$ | 0.37 | 49,420 | $4.50 \times 10^7$ |
| 3 | 17,958 | $1.25 \times 10^{-2}$ | $9.81 \times 10^{-13}$ | $8.78 \times 10^8$ | $1.85 \times 10^3$ | 3.03 | 18,616 | $4.15 \times 10^7$ |

**Table 1: Parameters characterizing the linear and nonlinear response of the vibrational modes**

## Supplementary Note 3: Shift in Resonance Frequency Due to Self- and Cross-Nonlinearity

In this section, we derive the resonance frequency shift of a given mode in the presence of both self-nonlinearity and cross-mode nonlinear coupling. We consider a system of two coupled nonlinear oscillators described by the following equations of motion:

$$\ddot{q}_i + 2\Gamma_i \dot{q}_i + \omega_{0,i}^2 q_i + \alpha_i q_i^3 + \kappa_{ij} q_j^2 q_i = m_i^{-1} F_d \cos(\omega_d t) \tag{16}$$

$$\ddot{q}_j + 2\Gamma_j \dot{q}_j + \omega_{0,j}^2 q_j + \alpha_j q_j^3 + \kappa_{ji} q_i^2 q_j = m_j^{-1} F_p \cos(\omega_p t) \tag{17}$$

where $q_i$ and $q_j$ denote the generalized displacements of the two modes, $\omega_{0,i}$ and $\omega_{0,j}$ are their respective natural frequencies, $\Gamma_i$ and $\Gamma_j$ are damping rates, $\alpha_i$ and $\alpha_j$ are the self-nonlinearity coefficients, and quantify the cross-mode coupling.

Assuming harmonic motion,

$$q_i(t) = A_i \cos(\omega_d t), \; q_j(t) = A_j \cos(\omega_p t), \tag{18}$$

the leading-order nonlinear contributions at frequency $\omega_d$ are:

$$q_i^3 = \frac{3}{4} A_i^3 \cos(\omega_d t), \quad q_i q_j^2 = A_i A_j^2 \cos(\omega_d t) \tag{19}$$

Substituting into the equation for $q_i$, neglecting damping and higher harmonics, the amplitude response becomes:

$$A_i = \frac{m_i^{-1} F_d}{\omega_{0,i}^2 - \omega_d^2 + \alpha_i \frac{3}{4} A_i^2 + \kappa_{ij} \frac{1}{2} A_j^2} \tag{20}$$

Setting the denominator to zero yields the resonance condition:

$$\omega_d^2 = \omega_{0,i}^2 + \frac{3}{4} \alpha_i A_i^2 + \frac{1}{2} \kappa_{ij} A_j^2 \tag{21}$$

Taking the square root and expanding to leading order gives:

$$\omega_d \approx \omega_{0,i} + \alpha_i \frac{3}{8} A_i^2 + \kappa_{ij} \frac{1}{4} A_j^2 \tag{22}$$

This result shows that the resonance frequency of mode $i$ is shifted by an amount proportional to both its own squared amplitude and that of the coupled mode $j$, due to self- and cross-mode nonlinearities, respectively.

## Supplementary Note 4: Parametric Modulation Due to Cross-Nonlinearity

We consider how nonlinear intermodal coupling leads to parametric modulation of the resonance frequency of a driven fundamental mode (mode $i$) when a higher-order mode (mode $j$) is excited by two nearby frequency components. This situation arises in the system described by Eqs. (16)-(17), where mode $i$ is directly driven and mode $j$ is perturbed by a near-resonant probe.

We assume the displacements take the form:

$$q_i(t) = A_i \cos(\omega_d t), \quad q_j(t) = A_{0,\text{eff},j} \cos(\omega_{0,\text{eff},j} t) + A_p \cos(\omega_p t) \tag{23}$$

where $A_{0,\text{eff},j}$ and $A_p$ are the amplitudes of mode $j$ at its resonance and at the probe frequency, respectively.

The nonlinear coupling term $q_j^2 q_i$ in the equation of motion for mode $i$ contains a beating component at frequency $\delta \omega_j = \omega_p - \omega_{0,\text{eff},j}$, leading to a time-dependent modulation of the effective stiffness of mode $i$. The resulting parametric modulation of its resonance frequency is:

$$\omega_{0,\text{eff},i}^2(t) = \omega_{0,i}^2 + \frac{3}{4} \alpha_i A_i^2 + \frac{1}{2} \kappa_{ij} A_j^2 + \kappa_{ji} A_{0,\text{eff},j} A_p \cos(\delta \omega_j t) \tag{24}$$

We define the static, nonlinearly shifted resonance frequency as:

$$\omega_{0,\text{shifted},i}^2 = \omega_{0,i}^2 + \frac{3}{4} \alpha_i A_i^2 + \frac{1}{2} \kappa_{ij} A_j^2 \tag{25}$$

Thus, the time dependent modulation becomes:

$$\omega_{0,\text{eff},i}^2(t) = \omega_{0,\text{shifted},i}^2 \left[1 + \delta \cos(\delta \omega_j t)\right] \tag{26}$$

With modulation depth:

$$\delta = \frac{\kappa_{ji} A_{0,j} A_p}{\omega_{0,\text{shifted},i}^2} \tag{27}$$

This analysis shows that nonlinear cross-mode coupling leads to parametric modulation of the fundamental mode's resonance frequency, driven by the spectral beating in the higher-order mode. In turn, this modulation can induce escape of the fundamental mode from its high-amplitude basin of attraction, as demonstrated in Fig. 5.

## Supplementary Note 5: Coupled Duffing modes in the rotating frame

We consider again the coupled nonlinear equations of motion for the two modes, Eqs. (16)-(17), and study the slow envelope dynamics of the fundamental mode $i$. Repeating similar procedures as in Supplementary Note 1, we obtain the quadratures equations in a rotating frame at the drive frequency $\omega_d$:

$$\dot{X}_i(t) = -\Gamma_i X_i(t) - Y_i(t)\left((\omega_i - \omega_{0,i}) - \frac{3\alpha_i}{8\omega_i}A_i^2(t) + \frac{\kappa_{ij}}{4\omega_i}A_j^2(t)\right) + \Im\{\bar{\xi}_{y,i}\},$$

$$\dot{Y}_i(t) = -\Gamma_i Y_i(t) + X_i(t)\left((\omega_i - \omega_{0,i}) - \frac{3\alpha_i}{8\omega_i}A_i^2(t) - \frac{\kappa_{ij}}{4\omega_i}A_j^2(t)\right) \tag{28}$$

$$+ \frac{F_i}{2m_i\omega_i}(1 + r\cos(\delta\omega t)) + \Re\{\bar{\xi}_{y,i}\},$$

where $X_{i,j}$ and $Y_{i,j}$ are the slowly varying quadratures of modes $i$ and $j$, respectively, and we have defined the instantaneous amplitudes as $A_{i,j}^2(t) = X_{i,j}^2(t) + Y_{i,j}^2(t)$. From these equations, we see that the nonlinear coupling terms $\kappa_{ij}A_j^2(t)$ modifies the effective resonance frequency of mode $i$ in the rotating frame. This cross-nonlinear contribution acts analogously to the self-Duffing shift, but arises from the dynamics of the coupled mode $j$, enabling frequency shifts that depend on the excitation of the higher-order mode.

## Supplementary Note 6: Instantaneous energy exchange rate of nonlinearly coupled modes

We derive the instantaneous energy transfer rate between two nonlinearly coupled vibrational modes. Starting from the equations of motion, this generalizes the results for linearly coupled harmonic oscillators and provides a physically intuitive criterion for identifying coupling regimes based on the rate of energy exchange.

We consider two modes with generalized coordinates $q_i(t)$ and $q_j(t)$, governed by:

$$\begin{aligned}\ddot{q}_i + 2\Gamma_i\dot{q}_i + \omega_{0,i}^2 q_i + \alpha_i q_i^3 + f_{ij}(q_i, q_i) &= m_i^{-1} F_i(t) \\ \ddot{q}_j + 2\Gamma_j\dot{q}_j + \omega_{0,j}^2 q_j + \alpha_j q_j^3 + f_{ji}(q_i, q_j) &= m_j^{-1} F_j(t)\end{aligned} \tag{29}$$

Here, $\omega_{0,i}$, $\Gamma_i$ and $\alpha_i$ denote the resonance frequency, and damping rate and Duffing nonlinearity coefficient of mode $i$, respectively. The coupling forces are represented by the functions $f_{ij}(q_1, q_2)$, with the total force exerted by mode $j$ on mode $i$ given by:

$$F_{j\to i} = m_i f_{ij}(q_i, q_j) \tag{30}$$

For reference, in the case of linear coupling, these interactions typically take the form:

$$f_{ij}(q_i, q_j) = gq_j$$
$$f_{ji}(q_i, q_j) = -gq_i \tag{31}$$

In the nonlinear case considered here, the interactions are:

$$f_{ij}(q_i, q_j) = \kappa_{ij} q_i q_j^2$$
$$f_{ji}(q_i, q_j) = \kappa_{ji} q_j q_i^2 \tag{32}$$

The corresponding instantaneous power transferred from mode $j$ to mode $i$:

$$P_{j \to i}(t) = F_{j \to i}(t) \dot{q}_i(t) = m_i f_{ij}(q_i, q_j) \dot{q}_i(t) \tag{33}$$

We define the instantaneous energy transfer rate from mode $j$ to mode $i$ as:

$$\Gamma_{j \to i}(t) = \frac{P_{j \to i}(t)}{\langle E_{\text{tot}} \rangle} \tag{34}$$

Where $\langle E_{\text{tot}} \rangle = \langle E_i \rangle + \langle E_j \rangle$ is the time-averaged total energy of both modes. This definition captures the real-time coupling dynamics relative to the characteristic energy scale of the system. Assuming harmonic motion:

$$q_i(t) = A_i \cos(\omega_{0,i} t + \phi_i)$$
$$\dot{q}_i(t) = -A_i \omega_{0,i} \cos(\omega_{0,i} t + \phi_i) \tag{35}$$

The time-averaged energy in each mode is:

$$\langle E_i \rangle = \frac{1}{2} m_i \omega_{0,i}^2 A_i^2 + \frac{1}{4} \alpha_i A_i^4 \tag{36}$$

Where the first and the second term account for the energy stored in the harmonic and anharmonic part of the potential, respectively. Although the harmonic contribution is dominant in our system, we include both terms in the full calculation for completeness.

In the linear case, assuming equal amplitudes, masses, and resonance frequencies ($\omega_{0,i} = \omega_{0,j} = \omega_0$, $A_i = A_j$, $m_i = m_j$), we find:

$$\Gamma_{i \leftrightarrow j}(t) = -\frac{g}{2\omega_0} \sin(2\omega_0 t). \tag{37}$$

The magnitude of the instantaneous energy transfer rate is amplitude-independent and the standard result for linearly coupled oscillators:

$$\Gamma_{i \leftrightarrow j}(t) = \frac{g}{2\omega_0}. \tag{38}$$

For the nonlinear coupling defined in Eq. (32), the instantaneous energy transfer rates become:

$$\Gamma_{j \to i}(t) = \frac{m_i \omega_{0,i} A_i^2 A_j^2}{4(\langle E_i \rangle + \langle E_j \rangle)} \left[ \sin(2\omega_{0,i} t) + \frac{1}{2} \sin(2(\omega_{0,i} + \omega_{0,j})t) + \frac{1}{2} \sin(2(\omega_{0,i} - \omega_{0,j})t) \right],$$

$$\Gamma_{j \to i}(t) = \frac{m_i \omega_{0,i} A_i^2 A_j^2}{4(\langle E_i \rangle + \langle E_j \rangle)} \left[ \sin(2\omega_{0,j} t) + \frac{1}{2} \sin(2(\omega_{0,j} + \omega_{0,i})t) + \frac{1}{2} \sin(2(\omega_{0,j} - \omega_{0,i})t) \right]. \tag{39}$$

Considering the dominant spectral component (i.e., with the largest amplitude) only, we obtain the magnitude of the instantaneous energy transfer rates

$$\Gamma_{j \to i} = \frac{\kappa_{ij} \omega_{0,i} m_i A_i^2 A_j^2}{4(\langle E_i \rangle + \langle E_j \rangle)} \tag{40}$$

$$\Gamma_{i \to j} = \frac{m_j \kappa_{ji} \omega_{0,j} A_i^2 A_j^2}{4(\langle E_i \rangle + \langle E_j \rangle)}$$

These rates are amplitude-dependent and generally asymmetric.

To classify the coupling regime, we compare the instantaneous energy transfer rates to the intrinsic dissipation rates of the modes. If

$$\Gamma_{i \to j}, \Gamma_{j \to i} > \max(\Gamma_i, \Gamma_j), \tag{41}$$

then energy is instantaneously exchanged faster than it dissipates, analogous to strong coupling. However if

$$\Gamma_{i \to j}, \Gamma_{j \to i} < \min(\Gamma_i, \Gamma_j) \tag{42}$$

Then dissipation will dominate before significant energy exchange occurs, analogous to weak coupling.